\documentclass[aps,prl,twocolumn,showpacs,preprintnumbers,amsmath,amssymb,superscriptaddress]{revtex4}
\usepackage{mathptm}  
\usepackage{dcolumn}                    
\usepackage{bm}                        
\usepackage{graphicx}
\usepackage{times}
\usepackage{epstopdf}
\begin{document}

\newcommand{\Imag}{{\Im\mathrm{m}}}   
\newcommand{\Real}{{\mathrm{Re}}}   
\newcommand{\talpha}{\tilde{\alpha}}
\newcommand{\im}{\mathrm{i}}        

\newcommand{\x}{\lambda}  
\newcommand{\y}{\rho}     
\newcommand{\T}{\mathrm{T}}   

\newcommand{\ml}{\boldsymbol{m}_L} 
\newcommand{\mr}{\boldsymbol{m}_R} 
\newcommand{\e}[1]{\mathrm{e}^{#1}}

\newcommand{\vecr}{\boldsymbol{r}} 
\newcommand{\ve}{\boldsymbol{E}} 
\newcommand{\vv}{\boldsymbol{v}}
\newcommand{\vb}{\boldsymbol{B}}
\newcommand{\vs}{\boldsymbol{S}}
\newcommand{\vk}{\boldsymbol{k}}
\newcommand{\vn}{\boldsymbol{\hat{n}}}

\newcommand{\eq}{Eq.}
\newcommand{\eqs}{Eqs.}
\newcommand{\cf}{\textit{cf. }}
\newcommand{\ie}{\textit{i.e. }}
\newcommand{\eg}{\textit{e.g. }}
\newcommand{\etal}{\emph{et al.}}
\def\i{\mathrm{i}}

\title{Spin-current in generic hybrid structures due to interfacial spin-orbit scattering}

\author{Jacob Linder}
\affiliation{Department of Physics, Norwegian University of
Science and Technology, N-7491 Trondheim, Norway}

\author{Takehito Yokoyama}
\affiliation{Department of Physics, Tokyo Institute of Technology, 2-12-1 Ookayama, Meguro-ku, Tokyo 152-8551, Japan}

\date{\today}

\begin{abstract}
We demonstrate a general principle that hybrid structures of any sort inevitably will give rise to a pure spin-current flowing parallel to the interface region when a charge-current is injected. This stems from the broken mirror symmetry near the interface which gives rise to spin-orbit coupling that deflects incoming electrons in a spin-discriminating fashion. We establish a general analytical condition for the appearance of this effect, and calculate the transverse spin-current explicitly using two different models. In addition, we investigate how the process of Andreev-reflection influences this phenomenon in the scenario where one of the materials is superconducting.  

\end{abstract}
\maketitle

Whereas the charge of electrons conventionally has been utilized in the field of electronics, the transport of electron spin and its belonging detection and manipulation has developed into a major research field in modern condensed matter physics over the last two decades \cite{zutic_rmp_04}. Besides its allure from a fundamental physics point of view, practical applications related to spintronics are already well-established \eg in the commercial harddisk-drive industry. More recently, state-of-the-art spintronics revolves around manipulation of domain walls in nanowires which could lead to the development of so-called race track memory devices \cite{parkin_science_08}. 

A central concept in the field of spintronics is the \textit{spin-current}: a flow of spin angular momentum carried by electrons. It is desirable to find experimentally feasible methods to generate spin-currents which are tunable and possible to exert control over. 
In fact, the spin-current is a key part of understanding phenomena such as the giant magnetoresistance effect \cite{baibich_prl_88, binash_prb_89} and current-induced spin-transfer torque \cite{slonczewski_jmmm_96, berger_prb_96}. 
One of the main challenges in spintronics is to create pure spin-currents for the purpose of spin injection. There exists several proposals which revolve around the spin Hall effect, spin-current generation transverse to the applied electric field due to spin-orbit coupling, to achieve this goal \cite{dyakonov_71, hirsch_prl_99,Murakami,Sinova}.

The bulk properties of a material are of high importance with regard to how spin-polarized transport may be obtained, often necessitating the use of materials with intrinsic ferromagnetic correlations in order to spin-polarize a current. However, the interface properties in hybrid structures are also an essential ingredient in this endevour. In fact, a generic property of any hybrid structure is that inversion symmetry is broken near the interface: the system is no longer centrosymmetric. As a direct consequence of this, a gradient of the electric potential will inevitably give rise to antisymmetric spin-orbit coupling effects for electrons moving in this region. This observation suggests a striking effect. In this Letter, we will demonstrate that the broken centrosymmetricity inherit to all hybrid structures invariably will generate a pure spin-current flowing parallel to the interface even in the absence of any magnetic elements or correlations, upon injection of charge-current (\eg by current- or voltage-bias). 

\begin{figure}[t!]
\centering
\resizebox{0.3\textwidth}{!}{
\includegraphics{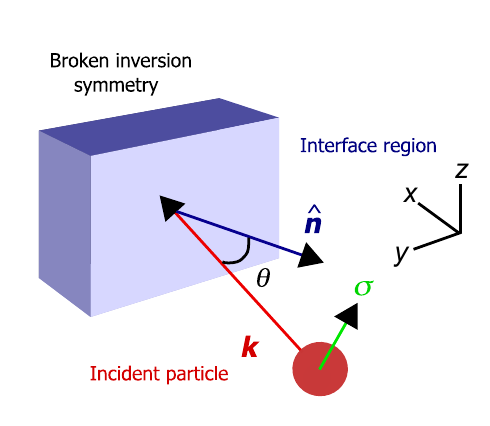}}
\caption{(Color online) Basic mechanism for the generic spin-current in hybrid structures. Centrosymmetricity is broken near the interface region, giving rise to an effective electric field. Incoming electrons thus feel an effective $\vk$-dependent magnetic field which couples to their spin. In this way, a pure transverse spin-current is generated near the interface region. In the calculations, we restrict our attention to a planar ($xy$) structure to obtain more transparent analytical results, although they may be generalized to 3D structures. }
\label{fig:model} 
\end{figure}

Let us explain the physical mechanism which gives rise to the general appearance of a transverse pure spin-current in a hybrid structure. Consider the model shown in Fig. \ref{fig:model}: a bilayer setup with two different materials separated by an interface region. Near the interface, mirror symmetry is broken and the system is non-centrosymmetric. As a result, an incoming electron with velocity $\vv$ experiences a net electric field $\ve$ due to the asymmetric crystal-field potential, induced perpendicularly to the interface. In the rest-frame of the electron, an effective magnetic field $\vb \propto \vv \times \ve$ appears. In turn, this gives rise to a spin-orbit coupling interaction of the form $\vb \cdot \vs \propto \vn\cdot(\vk \times \underline{\boldsymbol{\sigma}})$, where $\vn$ is an interface normal vector. As a consequence, the total interface potential matrix $\underline{V} = \underline{V_0} + \underline{V_\text{soc}}$ in spin space may be written as a sum of non-magnetic part ($V_0$) and spin-orbit coupling contribution ($V_\text{soc}$). 
This interface potential breaks spin symmetry while a transverse charge current is absent due to time-reversal symmetry. Therefore, the injected current generates \textit{a pure transverse spin-current} as a result of the broken centrosymmetricity, which is a \textit{generic feature of any hybrid structure.}

To illustrate this explicitly, we consider the model shown in Fig. \ref{fig:model} such that $\vn \parallel \boldsymbol{\hat{x}}$ and demonstrate that a pure spin-current will flow parallel to the interface, whereas there is no net spin transport across the junction. 
We will show that this transverse spin-current arises solely due to the broken centrosymmetricity, without any magnetic elements incorporated. To begin with, we analyze a prototype system consisting of a bilayer junction with two normal metals. We assume that there is a small region near the interface where spin-orbit coupling effects are present due to the lack of inversion symmetry. This barrier region is taken to have a finite width $L$ and a potential $U_0$ in addition to spin-orbit coupling $U_\text{soc}$. In this case, we may write: $\underline{V} = [U_0\underline{1} + U_\text{soc} \vn\cdot(\underline{\boldsymbol{\sigma}} \times \vk)/k_F]\Theta(x)\Theta(x-L),$
where $\Theta(x)$ is the Heaviside step-function. The wavefunction in the left normal metal region, where we will evaluate the generated spin-current, reads: $\psi_L = (\delta_{\sigma\uparrow}, \delta_{\sigma\downarrow})\e{\i qx} + r_\uparrow^\sigma(1,0)\e{-\i qx} + r_\downarrow^\sigma (0,1)\e{-\i qx}$. Here, $q = k_F\cos\theta$ is the wavevector in the normal region and $k_F$ is the magnitude of the Fermi momentum. The boundary conditions then demand $\psi_L = \psi_M$ and $\underline{v}_x^L\psi_L = \underline{v}_x^M\psi_M$ at $x=0$, whereas $\psi_R = \psi_M$ and $\underline{v}_x^R\psi_R = \underline{v}_x^M\psi_M$ at $x=L$. Here, $\psi_M$ and $\psi_R $ are wavefunctions in the middle and right regions, respectively. The velocity operators are given by $\underline{v}_x = \partial \underline{H}/\partial k_x$. The above set of equations may be solved analytically to provide the scattering coefficients of the system, which in turn can be used to calculate the spin-currents in the system. The spin-current is in general a tensor, with a direction of flow in real space and a polarization in spin space. For concreteness, we consider the $z$-component of the polarization and focus on its flow in the $x$- and $y$-directions which correspond to perpendicular and parallel to the barrier, respectively. The general definition reads:
\begin{align}\label{eq:spincurrent}
\mathcal{J}_{x,y} = \int^{\pi/2}_{-\pi/2} \text{d}\theta \sum_\sigma \frac{1}{2m} \text{Im}\{\psi^\dag \partial_{x,y} \underline{\sigma_z}\psi \}.
\end{align}
Inserting the wavefunction $\psi_L$ into Eq. (\ref{eq:spincurrent}), we note that $\mathcal{J}_x$ can be written generally as \footnote{We only consider transport near the Fermi level since our results of spin-current are obtaied as a linear response to a small applied bias. All transverse modes are taken into account.}: $\mathcal{J}_x = \int^{\pi/2}_0 \text{d}\theta k_F\cos\theta/(2m)\sum_{\alpha\sigma} [|r_\uparrow^\sigma(\alpha\theta)|^2 - |r_\downarrow^\sigma(\alpha\theta)|^2]$. Therefore, the general condition for the absence of a spin-current flowing across the barrier reads:
\begin{align}\label{eq:cond1}
\sum_{\alpha\sigma} [|r_\uparrow^\sigma(\alpha\theta)|^2 - |r_\downarrow^\sigma(\alpha\theta)|^2] = 0 \text{ with } \alpha=\pm1,\; \sigma =\uparrow,\downarrow.
\end{align}

\begin{figure}[t!]
\centering
\resizebox{0.4\textwidth}{!}{
\includegraphics{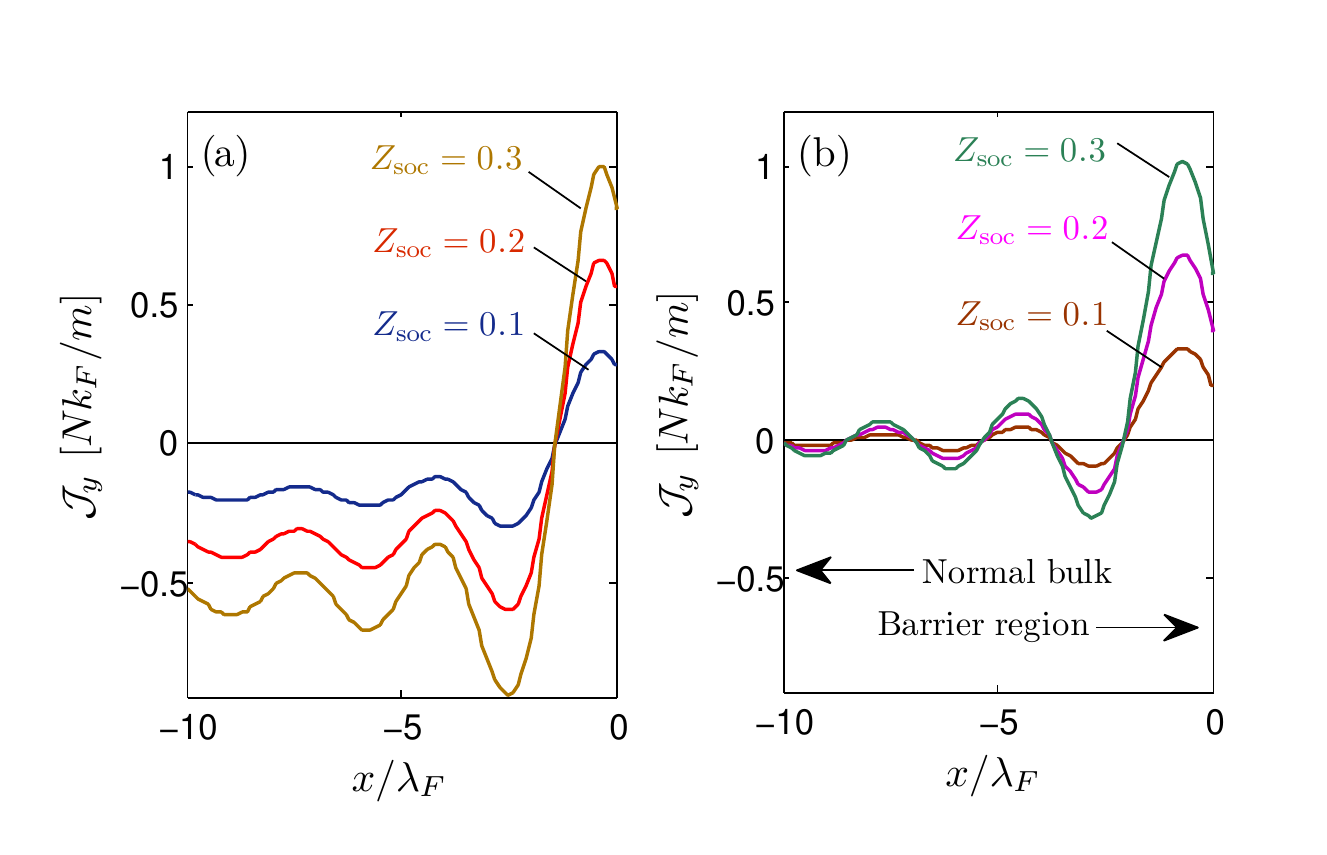}}
\caption{(Color online) Plot of the transverse spin-current $\mathcal{J}_y$ appearing in a normal metal$\mid$normal metal bilayer due to interface spin-orbit scattering. We show results for a barrier width of (a) $L=0.1$ nm and (b) $L=2$ nm, using in both cases a strong barrier potential $Z_0=5$.}
\label{fig:spincurrentNN} 
\end{figure}

Above, $\sigma$ corresponds to the spin of the incident particle whereas $\alpha$ is a sign index which arises when solving for the integral over $\theta$. Inserting this equation into the expression for $\mathcal{J}_y$, one arrives at: $\mathcal{J}_y = \int^{\pi/2}_0 \text{d}\theta (k_F \sin\theta/m)\sum_\sigma \text{Re}\{|r_\uparrow^\sigma(\theta)|^2 - |r_\downarrow^\sigma(\theta)|^2 + \e{-2\i qx}\sum_\alpha\sigma\alpha r_\sigma^\sigma(\alpha\theta)\}.$
This expression vanishes identically only when the following condition is satisfied:
\begin{align}\label{eq:cond2}
r_\uparrow^\sigma(\theta) = r_\downarrow^{-\sigma}(\theta)
\end{align}
If Eq. (\ref{eq:cond2}) is not satisfied, i.e. $r_\uparrow^\sigma \neq r_\downarrow^{-\sigma}$, the transverse spin-current remains finite. We underline here that both Eq. (\ref{eq:cond1}) and Eq. (\ref{eq:cond2}) are completely general expressions that have been derived without making any assumption about the barrier region and the details of the spin-orbit coupling present there. Let us now return to our specific model with a spin-orbit coupled barrier above contacted to two normal regions and use the potential $\underline{V}$. We then find the following specific expressions: $r_{\gamma}^\sigma = \delta_{\sigma\gamma}(q\rho^+_\gamma-k_x^\gamma \rho^-_\gamma)/(q\rho^+_\gamma+k_x^\gamma \rho^-_\gamma),$ $\rho^\pm_\gamma = 1 \pm \e{2\i k_x^\gamma L}(k_x^\gamma-q)/(k_x^\gamma+q),\; \gamma=\pm1=\uparrow,\downarrow$
for the reflection coefficients $r_\gamma^\sigma = \{r_\uparrow^\sigma, r_\downarrow^\sigma\}$. Above, $k_x^\gamma = \sqrt{2m(E_F + \gamma U_\text{soc}\sin\theta - U_0) - k_y^2}$ is the wavevector in the barrier region with $k_y = k_F\sin\theta$. Here, $U_\text{soc}$ and $U_0$ correspond to the spin-orbit coupling and barrier potential in the central region, respectively. We now prove explicitly that the result for $r_\gamma^\sigma$ dictates that \textit{the longitudinal spin-current vanishes whereas the transverse spin-current is finite}. By noting that $k_x^\gamma(\theta) = k_x^{-\gamma}(-\theta)$, we see that the condition $r_\uparrow^\sigma(\theta) = r_\downarrow^{-\sigma}(-\theta)$ is satisfied.
Since Eq. (\ref{eq:cond1}) is satisfied, it then follows that $\mathcal{J}_x=0$. However, Eq. (\ref{eq:cond2}) is not satisfied, and hence we have $\mathcal{J}_y\neq0$. Thus, we conclude that the presence of antisymmetric spin-orbit coupling at the interface due to the broken inversion symmetry induces a transverse spin-current whereas there is no spin flow across the barrier. The transverse spin-current and its dependence on the non-magnetic and spin-orbit coupling potentials is shown in Fig. \ref{fig:spincurrentNN}, with $\lambda_F=1/k_F$. For a normal metal with electron mass with approximately its bare value, $m \simeq m_0$, and a Fermi level of order eV, $E_F \sim 1$ eV, we obtain $k_F \simeq 5$ nm$^{-1}$. We show results for a barrier width of (a) $L=0.1$ nm and (b) $L=2$ nm, quantifying the potential in the barrier region by the dimensionless parameters $Z_0 = 2mU_0/k_F^2$ and $Z_\text{soc}=2mU_\text{soc}/k_F^2$. As seen, the spin-current increases in magnitude with $Z_\text{soc}$ and oscillates as a function of distance penetrated into the normal metal region. In both (a) and (b), we have set the atomic barrier potential to be strong, $Z_0=5$. The main difference between these two cases is that the spin-current decays to zero in the latter case whereas it saturates at a constant value in the former case. In order to understand this physically, we note that an increase of the barrier width $L$ inevitably generates less transmission through the bilayer such that the probability for reflection approaches unity. Examining the analytical expression for the transverse spin-current written above, it is clear that the constant terms ($x$-independent) cancel each other in this scenario, whereas the oscillatory term remains due to the different phases of the spin-$\uparrow$ and spin-$\downarrow$ reflection coefficients. Eq. (\ref{eq:cond1}) and (\ref{eq:cond2}) constitute a general criterion for the appearance of a pure transverse spin-current in a hybrid structure. Although there is no net spin-current flowing across the barrier, the interfacial spin-orbit scattering will invariably generate a transverse spin-current flowing parallel to the barrier according to the expression for $\mathcal{J}_y$. This spin-current will oscillate with the distance $|x|$ away from the barrier, and in general die out on a length-scale dictated by spin-relaxation processes due to \eg magnetic impurities and spin-flip scattering. 

\begin{figure}[t!]
\centering
\resizebox{0.4\textwidth}{!}{
\includegraphics{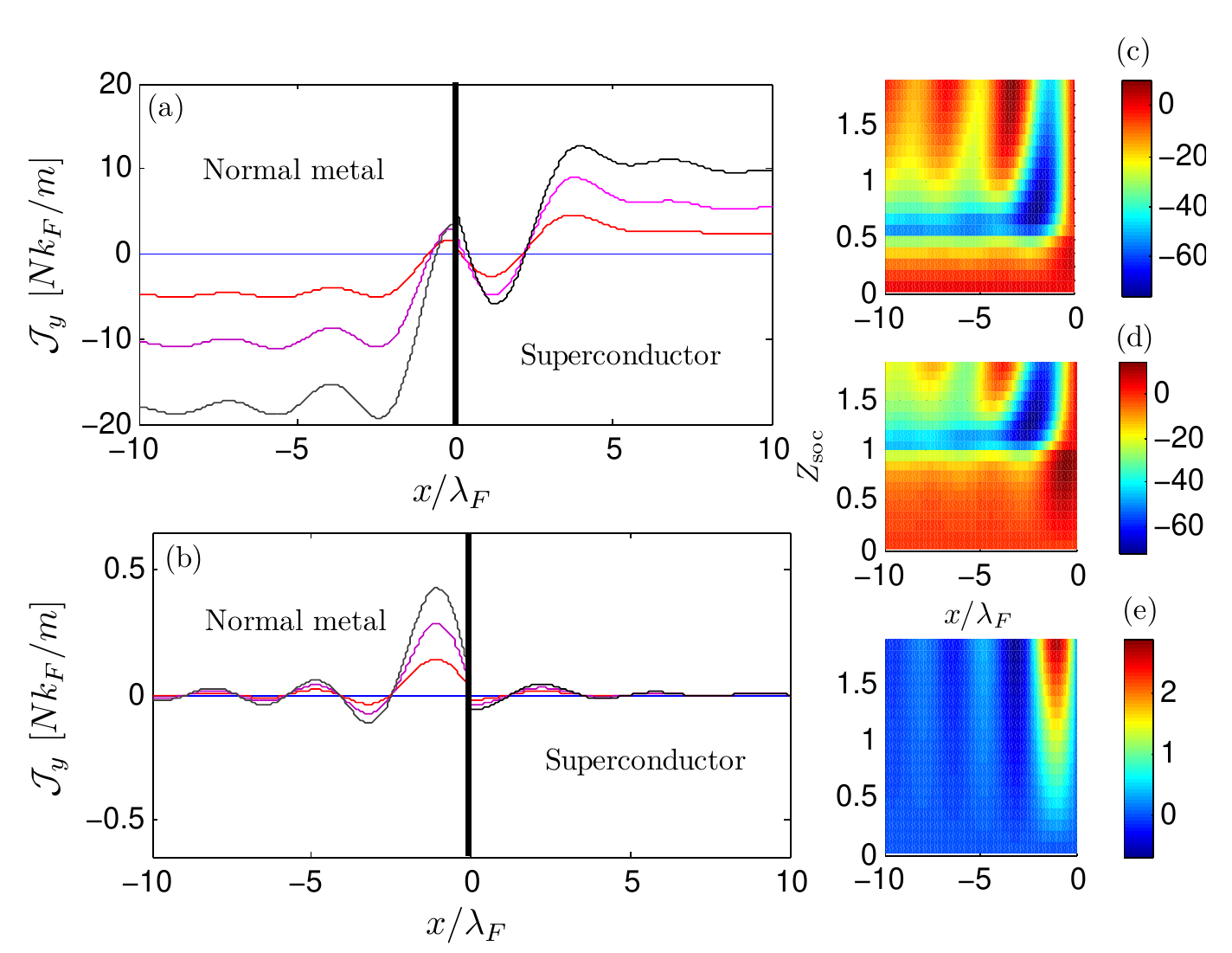}}
\caption{(Color online) Plot of the transverse spin-current $\mathcal{J}_y$ appearing in a normal metal$\mid$superconductor bilayer due to interface spin-orbit scattering. In (a) and (b), we plot the spin-current at Fermi level for $Z_\text{soc}=0.0$ to $0.3$ in steps of 0.1 for $Z_0=0.5$ and $Z_0=5$, respectively. In each panel, the curves increase in magnitude with $Z_\text{soc}$. In the side-panels, a contour-plot of the spin-current in the $x$-$Z_\text{soc}$ plane is given for the cases of (d) high $Z_0=0.5$, (e) intermediate $Z_0=1$, and (e) low $Z_0=5$ interface transparency.}
\label{fig:spincurrent} 
\end{figure}

Let us now consider an alternative model for the spin-orbit scattering at the interface region to demonstrate that the appearance of a transverse spin-current does not pertain to the specific model of a spin-orbit coupled barrier used above. Moreover, we turn our attention to the scenario where one of the contact materials is superconducting in order to see how the process of Andreev reflection influences the transverse spin-current. We employ the potential of the form \cite{wu_prb_09}: $\underline{V} = [V_0\underline{1} + V_\text{soc} \vn\cdot(\underline{\boldsymbol{\sigma}} \times \vk)/k_F]\delta(\vecr-\vecr_0),$
where $\vecr_0$ is the position of the interface, 
and consider for concreteness a bilayer consisting of a normal metal and an $s$-wave superconductor. This enables the electron-hole conversion by means of Andreev-reflection. Employing a scattering matrix formulation, we write down the incoming waves with spin $\sigma$ from the normal metal as: $\psi_N^\sigma = (1,0)\e{\i k_\theta x} + r_e^\sigma(1,0)\e{-\i k_\theta x} + r_A^\sigma(0,1)\e{\i k_\theta x}.$ In the superconductor, the transmitted wavefunction reads: $\psi_S^\sigma = t_e^\sigma(\e{\i\beta},\sigma)\e{\i k_\theta x} + t_h^\sigma(\sigma,\e{\i\beta})\e{-\i k_\theta x}$.
We incorporate any Fermi-vector mismatch as an effective increase in $V_0$. The boundary conditions at the interface location $x=0$ allow for a determination of the scattering coefficients $\{r_e^\sigma, r_A^\sigma, t_e^\sigma, t_h^\sigma\}$. These read: $\psi_S^\sigma = \psi_N^\sigma \text{ and } \partial_x(\psi_S^\sigma - \psi_N^\sigma) = 2m(V_0 - \sigma V_\text{soc} \sin\theta)\psi_N^\sigma.$
Upon defining $\mathcal{D}^\sigma = (Z^\sigma)^2 - \e{2\i\beta}[1+(Z^\sigma)^2]$ where $Z^\sigma = (Z_0/\cos\theta) - \sigma Z_\text{soc}\tan\theta$ and $Z_j = mV_j/k_F$, we obtain the following compact expressions for the scattering coefficients: $r_e^\sigma = Z^\sigma(\e{2\i\beta}-1)(\i + Z^\sigma)/\mathcal{D}^\sigma,$ $r_A^\sigma = -\sigma\e{\i\beta}/\mathcal{D}^\sigma,$
$t_e^\sigma = \e{\i\beta}(\i Z^\sigma-1)/\mathcal{D}^\sigma,$ $t_h^\sigma = -\i\sigma Z^\sigma/\mathcal{D}^\sigma.$
We emphasize that these equations \textit{are not simply a spin-polarized version of the Blonder-Tinkham-Klapwijk result} \cite{btk}. The interface spin-orbit coupling considered here gives rise to very different scattering coefficients than in the scenario of \eg a ferromagnet$\mid$superconductor bilayer or even a normal$\mid$superconductor layer with a magnetic interface where majority and minority spin carriers are shifted by an exchange field. 

Incorporating the possibility of having Andreev-reflection, one finds that an additional term $(-1)\times\int^{\pi/2}_0 \text{d}\theta k_F\sin\theta/(m)\sum_\sigma \sigma|r_A^\sigma(\theta)|^2$ is added to $\mathcal{J}_y$. Whereas this term vanishes when $Z_\text{soc}=0$, it remains finite in the presence of spin-orbit coupling and moreover enters with opposite sign to the probability for normal reflection in the expression for $\mathcal{J}_y$. Thus, whereas it is well-known that Andreev-reflection suppresses the longitudinal spin-current $\mathcal{J}_x$ in equal footing as normal reflection without branch-crossing, the Andreev-reflection probability enters with \textit{opposite sign} to the normal reflection for the transverse spin-current. The physical reason for this can be understood in terms of the retroreflective nature of the Andreev scattering process in the present system: the back-scattered hole has a group velocity opposite to its momentum, and hence propagates in the opposite direction of a normally reflected electron along the interface. Interestingly, it immediately follows that the Andreev-reflection coefficient in undoped graphene, where this scattering process can become specular \cite{beenakker_prl_06}, would contribute to the transverse spin-current in a different way than the present case. Thus, the transverse spin-current is actually sensitive to the exact nature of the Andreev-scattering process, which might be an interesting venue to explore further.

We are now in a position to again prove the existence of a non-zero transverse spin-current. The interface spin-orbit coupling does not violate time-reversal symmetry, and direct inspection of the above result shows that $r_e^\sigma(\theta) = r_e^{-\sigma}(-\theta)$ holds at the same time as $r_e^\sigma(\theta) \neq r_e^{-\sigma}(\theta)$ whenever $Z_\text{soc} \neq 0$. Based on our previous analysis, it follows that the transverse spin-current $\mathcal{J}_y \neq 0$. The inclusion of Andreev reflection alters the magnitude of the spin-current. We underline here that the superscript $\sigma$ for the reflection coefficient denotes both the spin of the incident particle and the spin of the reflected particle, such that $r^\sigma_e \equiv r^\sigma_{\sigma,e}$, which establishes the correspondence to Eqs. (\ref{eq:cond1}) and (\ref{eq:cond2}). It is also worth to note that the transmission coefficients satsify $t_e^\sigma(\theta) = t_e^{-\sigma}(-\theta)$ and $t_h^\sigma(\theta) = -t_h^{-\sigma}(-\theta)$ which can be analytically shown, leading to a non-zero spin current on the superconducting side near the interface.

Our treatment of the generated spin-current so far has been quite general as there are only two variational dimensionless parameters: the non-magnetic barrier potential $Z_0$ and the strength of the interface spin-orbit interaction $Z_\text{soc}$. We now plot the normalized spin-current $\mathcal{J}_y$ to investigate how it depends on these parameters, and also to see how it decays inside the bulk of the materials as one moves away from the interface. In Fig. \ref{fig:spincurrent}, we consider the behavior of $\mathcal{J}_y$ in an N$\mid$S bilayer for the case of a high (a) and low (b) interface transparency. In each case, several values of the spin-orbit scattering parameter $Z_\text{soc}$ are considered. As seen, the transverse spin-current persists in both the N and S region and increases in magnitude alongside with $Z_\text{soc}$. In each of the right subpanels of Fig. \ref{fig:spincurrent}, we provide a contour-plot of the spin-current to illustrate its behavior as a function of $Z_\text{soc}$ and position $x$ simultaneously. As seen, the spin-current experiences a sharp increase right around $Z_\text{soc} \simeq Z_0$, in effect where the spin-orbit scattering potential outgrows the non-magnetic barrier potential. 
We also find (not shown) that the spin-current is enhanced by the Andreev-reflection process compared to the case where the system is non-superconducting, $\Delta\to0$, when the interface transparency is high, i.e. low values of $Z_0$. With increasing barrier strength $Z_0$, the Andreev-reflection process is suppressed and the enhancement vanishes.

One should bear in mind that detection schemes for spin-currents generated by spin-orbit coupling are subject to constraints that arise from time-reversal symmetry in conjunction with the Onsager relations \cite{onsager}. For instance, it has been shown that whereas the detection of spins in a two-probe conductance setup cannot be accomplished by means of a magnetization switch, the detection of a torque signal generated by spin-flow is not prohibited by the symmetry \cite{adagideli_prl_06}.

Finally, we would like to comment on recent works which have considered how 
interfacial spin-orbit scattering arises. Whereas structural-inversion asymmetry (SIA) induced spin-orbit scattering traditionally has been discussed in the context of two-dimensional electron gases, \textit{ab initio} studies have also discussed this aspect for metallic surfaces \cite{henk_prb_03} and ferromagnet$\mid$semiconductor junctions (Fe$\mid$GeAs) \cite{gmitra_arxiv_09}. Moreover, a recent experiment demonstrated the spin-orbit scattering induced spin-torque by SIA in a ferromagnet$\mid$insulator junction (Co$\mid$AlO$_x$$\mid$Pt) \cite{miron_natmat_10} originally discussed in \cite{manchon_prb_08}. We believe that our model for the interfacial scattering should have merit since it has been shown that the surface spin-orbit scattering on \eg Au(111) \cite{henk_prb_03} is closely analogous to the Rashba interaction in two-dimensional electron gases. Having stated this, we also mention that the effect predicted in this Letter does not depend on the explicit Rashba-form used for the spin-orbit potential - it only depends on the asymmetry present for structural inversion. The analytical calculations that we have presented serve as a proof-of-principle for the generic effect of a transverse spin-current, whereas a quantitative estimate of the strength of the interfacial spin-orbit scattering and the exact plane of the bilayer junction which will support the asymmetric crystal-potential would be a task for \textit{ab initio} studies. 

In summary, we have theoretically demonstrated the appearance of transverse pure spin-current in
any hybrid structure, due to broken inversion symmetry near the interface.
We have provided the condition to realize this effect, and calculated the
spin-current for two bilayer systems. The predicted transverse spin-current
may be probed either via conversion into an electrical signal by the inverse
spin-Hall effect \cite{Saitoh,Valenzuela} or through optical studies such as
non-local Faraday rotation \cite{kikkawa_nature_99} and polarized
light-emission \cite{fiederling_nature_99}.

\end{document}